# Un modello di struttura dinamica per ebook scolastici.

Maria Vincelli

Università di Udine e Istituto di Linguistica Computazionale, CNR-Pisa

**Abstract**

Oggi non si possono ignorare le trasformazioni significative che le pratiche di studio e lo stile cognitivo delle nuove generazioni hanno subito a seguito dell'uso del Web [Paolo Ferri 2008; Angela Spinelli 2009]. La legge 133/2008 e le disposizioni ministeriali successive[1] concernenti i libri di testo per la scuola e relative all'obbligo del passaggio ai libri digitali o "misti", con tutti i limiti che si possono riscontrare in esse, rappresentano un riconoscimento non solo della necessità della transizione al digitale ma anche della sua positività culturale[2]. L'attuale offerta editoriale scolastica, in ottemperanza alla legge su citata, concepisce però il libro di testo digitale sostanzialmente come una trasposizione del libro cartaceo tradizionale, in formato PDF o ePub, e prevede dunque le medesime modalità di fruizione. I "libri misti" hanno permesso alle case editrici di potenziare i libri di testo (cartacei o digitali) corredandoli di tutta una serie variegata, ma spesso incoerente, di materiali e servizi integrativi "esterni" al libro stesso. In ogni caso, le funzionalità di consultazione e di lettura dei libri digitali scolastici risultano alquanto insoddisfacenti.

In questo panorama due iniziative si sono imposte recentemente come nuove e interessanti: "Create" della McGraw-Hill, e "Custom Publishing–Il libro fatto su misura", della Pearson[3]. Il servizio offerto ai docenti è quello di creare un libro su misura, componendo capitoli attinti da libri diversi, naturalmente dei propri cataloghi. Il vantaggio della personalizzazione di un libro di testo è di facile comprensione. Di contro, la frammentazione non controllata del materiale rappresenta un grave limite: non è detto, infatti, che fra le varie parti ricomposte ci sia coerenza nell'impostazione metodologica e nello stile. Inoltre rimane alto il grado di arbitrio nella scelta e nella strutturazione degli argomenti; ancora, alla fine del processo di scomposizione e ricomposizione, si avrà comunque un libro lineare e chiuso rispetto alla fruizione da parte degli studenti. Più in generale la frammentazione non controllata rischia di depauperare il libro di testo del valore culturale di cui esso è portatore, proprio in quanto espressione di alte competenze disciplinari e di un'unità di senso che ad esse viene impressa all'atto della composizione da parte dell'autore, nel momento in cui vengono ordinate in una sequenza lineare e all'interno di una struttura che le contiene.

---

[1] I riferimenti normativi principali sono la Legge 133/2008 e la 169/2008.
[2] Che sia auspicabile e necessaria una metamorfosi del libro tradizionale, non vuol dire di per sé eliminazione del libro di testo. A questo proposito, Paolo Ferri scrive: «Analizziamo per esempio come può essere inteso nell'epoca della società informazionale uno strumento antico e carico di gloria: il libro scolastico. I testi scolastici, che sono ancora oggi (e lo saranno per i prossimi anni) lo strumento per la mediazione educativa dei contenuti stanno subendo e subiranno sempre di più una metamorfosi digitale», in *La scuola digitale*, Bruno Mondadori 2008, p. 97.
[3] Questi servizi sono disponibili rispettivamente ai seguenti indirizzi: http://create.mcgraw-hill.com/wordpress-mu/unitedstates/ e http://hpe.pearson.it/docenti-custom-publishing.php.

In questo articolo si propone un modello di ebook per le scuole basato su una struttura a grafo in cui i nodi rappresentano singoli argomenti di una programmazione didattica a un livello piuttosto ridotto di granularità, che faciliti la loro aggregazione e riusabilità, e gli archi rappresentano legami di propedeuticità fra argomenti (e quindi, appunto, fra nodi).[4] Su questo grafo svilupperemo una serie di semplici algoritmi che permettono sia ai docenti sia agli studenti di ricomporre in maniera interattiva e personalizzata le varie unità disponibili secondo una successione che, rispettando le propedeuticità, risulterà alla fine consistente sotto l'aspetto metodologico e stilistico in senso lato. Pertanto docenti e studenti non avranno a disposizione un insieme scorrelato di unità né un insieme limitato di pochi percorsi didattici preconfezionati, come è tipico di alcune soluzioni sul Web, ma piuttosto disporranno di una rete di argomenti che potranno essere serializzati in un numero combinatoriamente molto vasto di alternative, e quindi potranno creare altrettanti ebook personalizzati ma garantiti sotto il profilo scientifico.

In questo articolo dimostreremo che, sorprendentemente, le potenzialità offerte da questo modello di ebook si estendono ben al di là della serializzazione controllata di una serie di argomenti, e riguardano un insieme di funzionalità, attualmente assenti negli ebook scolastici, del tutto auspicabili e in parte contemplate dal DL 41/2009. Per brevità illustreremo solo alcune di queste, quali la possibilità:

- Per i docenti: di gestire con efficacia le attività interdisiciplinari (per es. fra discipline dell'asse linguistico), prevedendo la possibilità di sincronizzare grafi elaborati per le singole discipline; di intervenire in maniera efficace e sistematica nel recupero di lacune dell'apprendimento con la creazione di ebook personalizzati di ripasso; di

---

[4] Si tratta di una forma di iper-link denominata lexia-to-lexia da George P. Landow in *Hypertext 3.0*, The Johns Hopkins University Press, 2006. In questo articolo proponiamo un uso innovativo di questo semplice modello nell'ambito degli ebook scolastici, che si avvarrà di algoritmi che ne esaltano le potenzialità didattiche, la personalizzazione e la fruibilità dei testi digitali.

- integrare nel modello materiale attinto da altre fonti, anche di natura multimediale e di personale elaborazione, in un insieme che possa essere coerente.
- Per gli studenti: di gestire efficacemente lo studio individuale, in particolare il ripasso per il consolidamento o il recupero, arrivando a disporre di un ebook realmente efficace, che si può focalizzare sugli argomenti di proprio interesse; o di utilizzare il grafo come mappa concettuale per collegare nuove e vecchie conoscenze, operazione considerata decisiva nella teoria dell'apprendimento significativo di cui Novak [2010] si fa portavoce.

Per quanto riguarda gli esercizi, il grafo può essere utilizzato secondo due modalità diverse. Sullo stile dei libri cartacei ogni nodo (argomento) ne avrà di propri; ma a differenza dei libri cartacei si potranno prevedere degli esercizi "esterni" a un singolo nodo, e in verità "associati" a più nodi cui si riferiscono in termini di competenze che lo studente deve possedere per poterli svolgere. In questo secondo caso, il libro personalizzato collocherà questi esercizi nella posizione in cui queste competenze saranno state acquisite dallo studente, secondo la serializzazione assunta dall'ebook. Un'altra interessante applicazione del grafo, specifica del caso di studio descritto successivamente, consiste nel consentire ai docenti e agli studenti di analizzare automaticamente parole o frasi (in latino) servendosi di strumenti sofisticati quali gli analizzatori morfologici (menziono fra gli altri LemLat, messo a punto dal CNR di Pisa, disponibile all'indirizzo http://www.ilc.cnr.it/lemlat/). Strumenti del genere possono infatti identificarne la forma grammaticale, e quindi sarebbe possibile per un software risalire alle unità/nodi corrispondenti nel grafo. A questo punto il grafo potrebbe essere visitato per indicare al docente/studente l'insieme delle competenze che sono necessarie per poter capire quella forma o svolgere quell'esercizio; o, equivalentemente, potrebbe essere utilizzato per "associare" quell'esercizio o quella forma a quell'insieme di unità per una eventuale visualizzazione futura, secondo la modalità di esercizi "esterni" indicata precedentemente.

Per dimostrare l'efficacia del modello proposto si è scelto come caso studio l'insegnamento della lingua latina al biennio delle superiori, e si è realizzato un grafo di circa 150 nodi e 300 archi utilizzando un software di nome yEd-Graph Editor (disponibile all'indirizzo www.yworks.com/products/**yed**/). Per la determinazione della granularità degli argomenti corrispondenti ai nodi è stato scelto un libro scolastico di lingua latina composto da Sebastiano Timpanaro, *De Lingua Latina*, pubblicato nel 1990 dalla casa editrice Liviana di Padova. Questo libro ha la peculiarità di presentare un'organizzazione complessiva del materiale ben adattabile alla scomposizione di cui si è detto sopra: la trattazione è suddivisa in 219 lezioni dedicate ad argomenti autoconsistenti che occupano mediamente circa 3,5 pagine, per un totale di 790 pagine. Per quanto riguarda gli archi del grafo, questi sono stati derivati analizzando il contenuto delle lezioni e identificando le necessarie propedeuticità tra esse. YEd offre anche la possibilità di effettuare delle visite del grafo, e quindi implementare gran parte delle funzionalità su descritte per verificarne la loro efficacia.

Si tratta evidentemente di un *proof-of-concept*, ciò nonostante la descrizione offerta da questo articolo dovrebbe convincere facilmente il lettore della fattibilità e dell'efficacia del modello di ebook scolastico proposto. Stiamo realizzando un prototipo software che implementa tutte le funzionalità descritte sul grafo di 150 nodi indicato precedentemente e con i contenuti estratti dal libro di Timpanaro. Questo prototipo dovrebbe essere disponibile per le date della conferenza.

Il lavoro descritto in questo articolo rientra nel progetto della tesi di dottorato in "Scienze bibliografiche, del testo e del documento" in corso di svolgimento presso l'Università di Udine sotto la supervisione del Prof. Guarasci (Università della Calabria) e del dott. Bozzi (Direttore dell'Istituto ILC CNR Pisa).

1. **Appendice**

Come indicato nell'abstract, per dimostrare l'efficacia del modello proposto si è scelto un caso studio che è l'insegnamento della lingua latina al biennio delle superiori, e si è realizzato un grafo di circa 150 nodi e 300 archi utilizzando un software di nome yEd-Graph Editor (disponibile all'indirizzo www.yworks.com/products/**yed**/). Per la determinazione della granularità degli argomenti corrispondenti ai nodi è stato scelto un libro scolastico di lingua latina composto da Sebastiano Timpanaro, *De Lingua Latina*, e pubblicato nel 1990 dalla casa editrice Liviana di Padova. Questo libro ha la peculiarità di presentare un'organizzazione complessiva del materiale ben adattabile a una scomposizione in unità di granularità ridotta, che faciliti quindi la loro aggregazione e riusabilità: la trattazione è suddivisa in 219 lezioni dedicate ad argomenti autoconsistenti che occupano mediamente 3,5 pagine, per un totale di 790 pagine. Per quanto riguarda gli archi del grafo, questi sono stati derivati analizzando il contenuto delle lezioni e identificando le necessarie propedeuticità tra esse.

Il grafo soddisfa i seguenti requisiti:

- Tutti gli archi sono orientati, per codificare le propedeuticità fra gli argomenti/nodi che connettono.
- Il grafo è aciclico, ossia le propeduticità sono tali da non vincolarsi in maniera circolare: perciò se l'argomento A è propedeutico a B, e B è propedeutico a C, non deve verificarsi che C sia propedeutico ad A.
- Da un nodo possono uscire più archi, e più archi possono entrare in un solo nodo (i nodi da cui escono più archi saranno quelli che corrispondono ad argomenti basilari, ad esempio alle categorie grammaticali fondamentali).

Per facilitare la descrizione del modello e le sue funzionalità, nel seguito si farà riferimento a un grafo ridotto, estratto dal precedente, costituito da 32 nodi e 33 archi. I nodi scelti corrispondono ragionevolmente al programma di Lingua latina di un primo trimestre.

Nella figura 1 è rappresentata la struttura a grafo che sarà soggiacente a un libro digitale di lingua latina. Nel grafo si osserva che tutti i nodi hanno archi in uscita orientati; che alcuni nodi non hanno archi in entrata (sono quelli corrispondenti ad argomenti più generali); che gli archi sono di colore diverso, per indicare diversi valori delle propedeuticità: il colore verde indica una relazione di propedeuticità opzionale; il nero indica una relazione di propedeuticità necessaria; il rosso indica una relazione di propedeuticità sempre necessaria ma alternativa a quella codificata con il nero. Un docente che voglia ritagliarsi un libro meglio rispondente alla sua programmazione potrà selezionare gli argomenti di suo interesse (ragionando auspicabilmente per obiettivi di apprendimento a medio-lungo termine) e derivare dal grafo anche tutti gli altri argomenti che sono ad essi propedeutici e quindi costituiscono i suoi "predecessori" nella rete definita dalla struttura. Questi argomenti sono quelli che devono sicuramente essere svolti prima di arrivare all'argomento evidenziato.

La figura 2 mostra come il sistema, a partire dagli argomenti selezionati dal docente (nell'esempio il docente specifica la "Proposizione causale"), effettua quella che si chiama "visita del grafo a ritroso" e quindi identifica un sottografo del grafo originario nel quale sono contenuti tutti i "predecessori" del nodo selezionato dal docente e le relazioni fra questi (la visita corrisponde alla parte del grafo evidenziata in verde). Una volta effettuata la "visita" si può procedere a quello che in gergo informatico si chiama "ordinamento topologico", vale a dire la linearizzazione dei nodi del sottografo in modo tale che tutti gli archi fra essi risultino orientati da sinistra a destra. La linearizzazione rappresenta la sequenza logico-temporale con cui quegli argomenti devono essere svolti, e quindi, offre una possibile codifica di libro di testo per quegli argomenti. Il libro così ottenuto è diverso da quello di partenza non solo perché contiene un sottoinsieme degli argomenti, ma anche perché la successione di questi può essere diversa da quella originale, sempre e comunque compatibile con le propedeuticità imposte dal grafo.

La figura 3 rappresenta alcune delle linearizzazioni compatibili con le propedeuticità definite nel grafo originario, e dimostra al contempo la potenza codificatrice del modello proposto rispetto alle possibilità offerte dalla singola sequenza, propria del libro cartaceo, o dall'insieme di percorsi rigidamente precodificati, tipica delle realizzazioni più diffuse sul Web. Di queste sequenze, comunque, non tutte sono ugualmente significative e rilevanti da un punto di vista didattico. L'obiettivo futuro della presente ricerca sarà quindi quello di definire dei criteri ragionevoli per effettuare un *ranking* di esse, offrendo così ai docenti e agli studenti dei suggerimenti tra le "migliori" possibili serializzazioni. Criteri interessanti potrebbero essere per esempio il tempo stimato per effettuare tutto il percorso; oppure una sorta di accreditamento, sulla base della quantità di accessi a ciascun percorso.

In conclusione osservo che le potenzialità offerte da questo modello di libro si estendono ben aldilà della serializzazione controllata di una serie di argomenti e riguardano un insieme di funzionalità, attualmente assenti negli ebook scolastici, del tutto auspicabili. Ne menziono di seguito solo alcune:

**per i docenti:**

- Personalizzare il libro sulla base della propria programmazione, e poterlo gestire modificandolo *in itinere*.
- Gestire con efficacia le attività interdisiciplinari (per es. fra discipline dell'asse linguistico), prevedendo la possibilità di sincronizzare grafi elaborati per le singole discipline.
- Intervenire in maniera efficace e sistematica nel recupero di lacune dell'apprendimento con la creazione di percorsi personalizzati di ripasso.

**per gli studenti:**

- Collegare nuove e vecchie conoscenze, operazione considerata decisiva nella teoria dell'apprendimento significativo di cui Novak si fa portavoce.
- Gestire efficacemente lo studio individuale, in particolare il ripasso per il consolidamento o il recupero, arrivando a disporre di un libro realmente efficace, che si focalizza sugli argomenti di proprio interesse.

Per quanto riguarda gli esercizi, il grafo può essere utilizzato secondo due modalità diverse. Sullo stile dei libri cartacei ogni nodo (argomento) ne avrà di propri; ma a differenza dei libri cartacei si potranno prevedere degli esercizi "esterni" a una singola unità, che sono in verità "associati" a più nodi/unità cui si riferiscono in termini di competenze che lo studente deve possedere per poterli svolgere. In questo secondo caso, il sistema potrà visualizzare questi esercizi soltanto nel momento in cui queste competenze saranno state acquisite dallo studente in accordo al libro-personalizzato che è stato generato. Quindi non una collocazione statica di questi esercizi, ma una collocazione che si adatta dinamicamente al libro-personalizzato secondo la sua forma che il docente/studente definisce.

Un'altra interessante applicazione del grafo consiste nella possibilità di supportare l'integrazione automatica nel libro di altro materiale esterno, a scopo di esercitazione/valutazione, servendosi di strumenti sofisticati quali gli analizzatori morfologici (menziono fra gli altri LemLat, messo a punto dal CNR di Pisa). Strumenti del genere possono infatti consentire ai docenti e agli studenti di analizzare automaticamente parole o frasi in latino, identificarne la forma grammaticale, e risalire così alle unità/nodi corrispondenti. A questo punto il grafo potrebbe essere visitato per indicare al docente/studente l'insieme delle competenze che sono necessarie per poter capire quella forma o svolgere quell'esercizio; o, equivalentemente, potrebbe essere utilizzato per "associare" quell'esercizio o quella forma a quell'insieme di unità per una eventuale visualizzazione futura, secondo la modalità di esercizi "esterni" indicata precedentemente.

## 2. Le criticità del modello proposto

Nel modello proposto si evidenziano delle criticità, che sono di seguito elencate in ordine decrescente di significatività:

1) La determinazione dell'unità minima corrispondente a un nodo è arbitraria.
2) Il rapporto di propedeuticità non è sempre oggettivo: mantiene in alcuni casi un margine di arbitrarietà, che in parte viene superato se la valutazione viene fatta in considerazione degli obiettivi a medio termine che il docente si propone.
3) Numero elevato di sequenze che serializzano il sottografo dei predecessori, e quindi necessità di definire delle funzioni di *ranking* che permettano di restringere l'attenzione del docente/studente a un sottoinsieme che meglio si adatta alle esigenze didattiche/apprendimento.

Per superare queste criticità, in particolare le prime due, bisogna innanzitutto tenere ben presente che il modello, una volta elaborato, deve riceve la validazione da parte di esperti della disciplina, sia in ambito universitario sia da parte degli stessi docenti, e la validazione non dovrà necessariamente portare a un modello che si critallizza, disciplina per disciplina, una volta per tutte. Il modello dovrà e potrà perfezionarsi, recependo i contributi più proficui dei dibattiti sullo statuto delle discipline e sulle metodologie didattiche. Inoltre, in fase di elaborazione del grafo (come si è fatto nel caso specifico del grafo proposto), bisognerà procedere a un lavoro scrupoloso di studio dei libri di testo migliori, selezionati per l'alto grado di affidabilità, e di confronto fra questi, tenendo sempre come riferimento fermo le trattazioni scientifiche che riguardano problematiche particolari dell'insegnamento delle singole discipline e delle diverse metodologie.

## 3. Conclusioni

Ci proponiamo di perfezionare il grafo e validarlo con il supporto di un gruppo di docenti (validazione già parzialmente compiuta verificando la compatibilità di diversi libri di testo di lingua latina con le possibili serializzazioni del grafo corrente); inoltre, si procederà a realizzare un prototipo che implementi le funzionalità precedentemente descritte per il modello di libro digitale su proposto: in questa realizzazione i nodi saranno "riempiti" con schede di contenuto attinte dal libro *De Lingua Latina* di Timpanaro, digitalizzate e opportunamente adattate allo scopo. Questo prototipo sarà pronto per le date della conferenza e verrà utilizzato per fornire un *proof-of-concept* dell'efficacia e delle funzionalità realizzabili con il modello di libro digitale proposto in questo articolo.

## Bibliografia

Ferri, Paolo. *La scuola digitale. Come le nuove tecnologie cambiano l'informazione*. Milano: Bruno Mondadori, 2008.

Landow, George. *Hypertext 3.0. Critical Theory and New Media in an Era of Globalization*. Baltimore: Johns Hopkins University Press, 2006.

Novak, Joseph. *Learning, Creating and Using Knowledge. Concept Maps as Facilitative Tools in Schools and Corporations*. New York: Routledge, 2010, 2^ ed.

Pasini, Aureliana, Timpanaro, Sebastiano. *De Lingua Latina. Corso di Latino per il biennio delle superiori*. Padova: Liviana editrice, 1990.

Roncaglia, Gino. *La quarta rivoluzione. Sei lezioni sul futuro del libro*. Roma-Bari: Laterza, 2010.

Spinelli, Angela. *Un'officina di uomini. La scuola del costruttivismo*. Napoli: Liguori, 2009.

# Figura 1

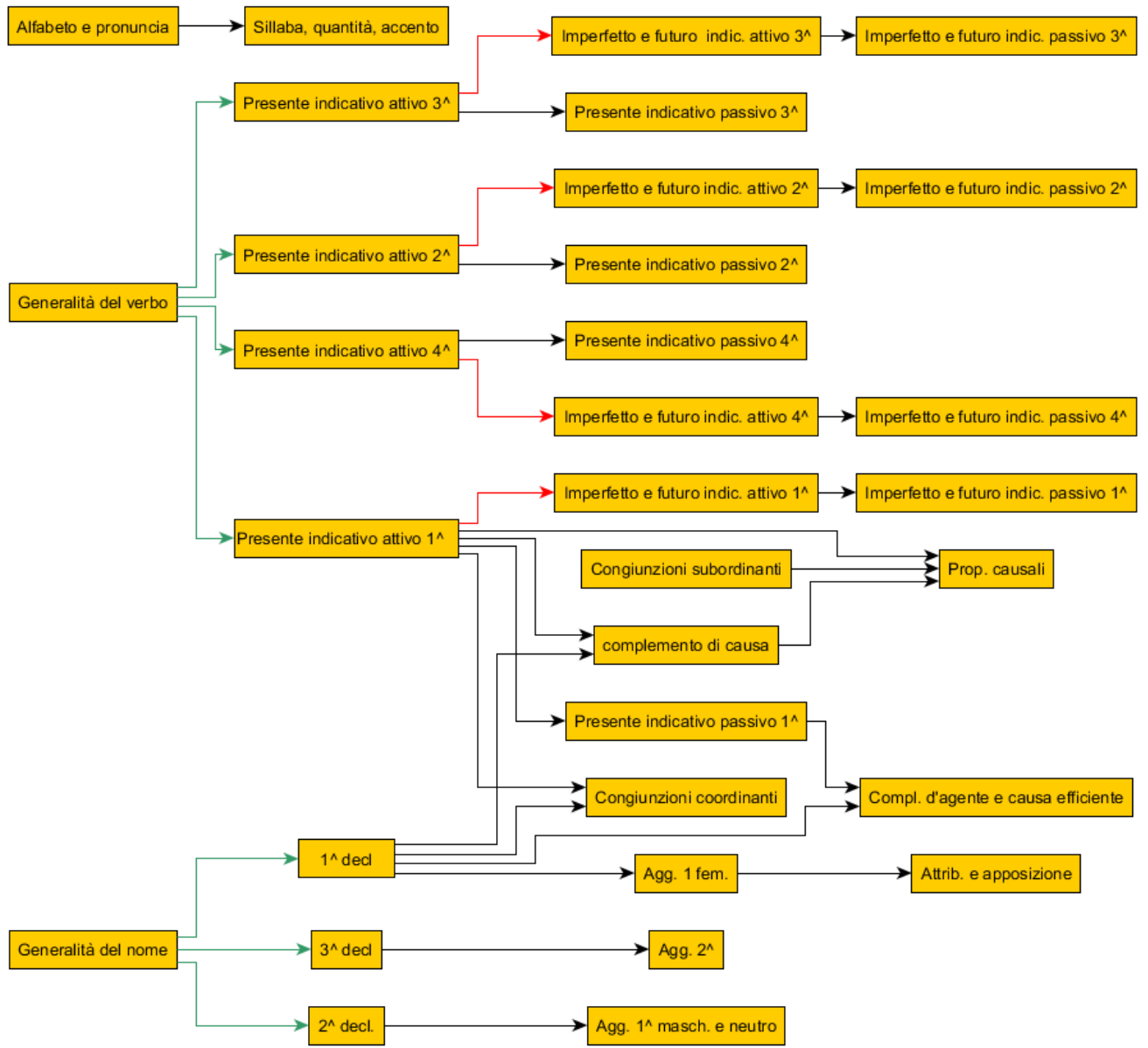

# Figura 2

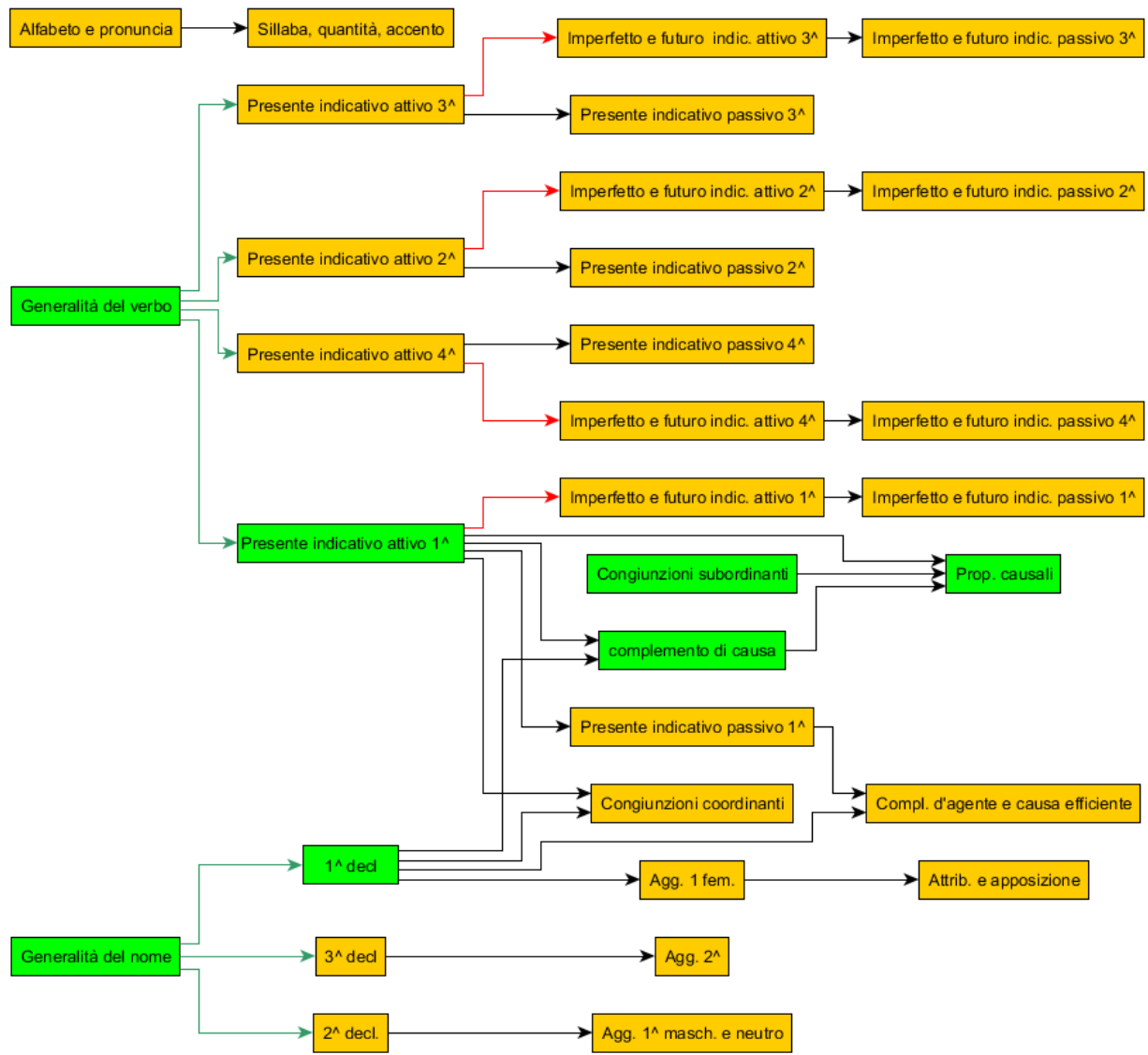

# Figura 3

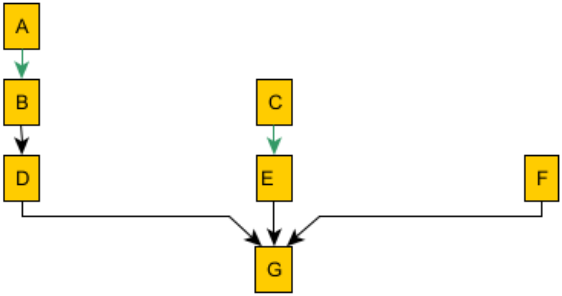
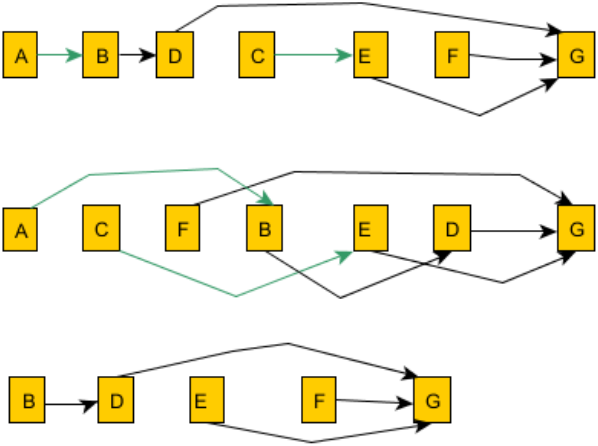